\documentclass[twocolumn,british,pra,superscriptaddress,showpacs]{revtex4-1}
\usepackage[letterpaper]{geometry}
\usepackage{graphicx}
\usepackage{epstopdf}
\usepackage{amsmath}
\usepackage{amssymb}
\usepackage{centernot}
\usepackage{float}
\newcommand{\bra}[1]    {\langle #1|}
\newcommand{\ket}[1]    {| #1 \rangle}

\renewcommand{\t}[1]{\textrm{#1}}
\newcommand{\tildevarphi}{\widetilde{\varphi}}
\newcommand{\mean}[1]   {\langle #1 \rangle}
\renewcommand{\d}       {\mathrm{d}}
\newcommand{\Tr}        {\mathrm{Tr}}

\renewcommand{\v}[1]    {\mathbf{#1}}

\newcommand{\ii}{\mathrm{i}}
\makeatletter
\@ifundefined{textcolor}{}
{%
 \definecolor{BLACK}{gray}{0}
 \definecolor{WHITE}{gray}{1}
 \definecolor{RED}{rgb}{1,0,0}
 \definecolor{GREEN}{rgb}{0,1,0}
 \definecolor{BLUE}{rgb}{0,0,1}
 \definecolor{CYAN}{cmyk}{1,0,0,0}
 \definecolor{MAGENTA}{cmyk}{0,1,0,0}
 \definecolor{YELLOW}{cmyk}{0,0,1,0}
 }

\makeatother

\makeatother

\usepackage{babel}

\begin{document}

\title{Usefulness of an enhanced Kitaev phase-estimation algorithm in quantum
metrology and computation}

\author{Tomasz Kaftal}
\affiliation{Institute of Theoretical Physics, University of Warsaw, ul. Ho\.{z}a 69, PL-00-681 Warszawa, Poland}

\author{Rafa{\l} Demkowicz-Dobrza{\'n}ski}
\affiliation{Institute of Theoretical Physics, University of Warsaw, ul. Ho\.{z}a 69, PL-00-681 Warszawa, Poland}

\begin{abstract}
We analyze the performance of a generalized Kitaev's phase estimation algorithm
where $N$ phase gates, acting on $M$ qubits prepared in a product state, may be
distributed in an arbitrary way. Unlike the standard algorithm, where the mean
square error scales as $1/N$, the optimal generalizations
offer the Heisenberg $1/N^2$ error scaling and we show that they are in fact
very close to the fundamental Bayesian estimation bound. We also demonstrate
that the optimality of the algorithm breaks down when losses are taken into
account, in which case the performance is inferior to the optimal
entanglement-based estimation strategies.  Finally, we show that when an
alternative resource quantification is adopted, which describes the phase
estimation in Shor's algorithm more accurately, the standard Kitaev's procedure
is indeed optimal and there is no need to consider its generalized version.
\end{abstract}


\maketitle

\section{Introduction}
Phase estimation is a key problem in many physical experiments involving quantum
phenomena and, as such, is an important research topic in quantum metrology
\cite{Giovannetti2006,Giovannetti2011}. Precise quantum-enhanced measurement of
the optical phase is the cornerstone of quantum interferometry and finds impressive
applications in, e.g., gravitational wave detection \cite{Abadi2011}. Limits on
phase-estimation precision in realistic scenarios have been determined using
fundamental concepts of quantum estimation theory, such as quantum Fisher
information \cite{Knysh2010, Escher2011, Demkowicz2012, Knysh2014} as well as Bayesian
inference \cite{Kolodynski2010}. They give valuable insight into how much
information can be extracted in such experiments. Phase-estimation protocols may
also be used as a subroutine in quantum algorithms. Most notably, Kitaev's
phase-estimation algorithm \cite{Kitaev1997} lies at the heart of the famous
Shor's algorithm \cite{Shor1997, Nielsen2000}.

In this article we combine the algorithmic and metrological approaches to phase
estimation by proposing a generalization of Kitaev's algorithm and showing that
it comes indiscernibly close to the Bayesian estimation optimum, even though no
entanglement is present in the probe states used. The algorithm can be realized
using single-photon multipass interferometric strategies similar to those
studied in \cite{Higgins2007} and \cite{Higgins2009}. Still, unlike
\cite{Higgins2007} and \cite{Higgins2009} we assume the most general
measurement-estimation
scheme and hence obtain significantly better estimation performance under
fewer available resources. Simultaneously we present a generic recipe for
analyzing the performance of Kitaev-like algorithms in the Bayesian approach. We
then show how it can be used to rederive results from \cite{Berry2009} on
the asymptotic  bounds on the performance of such algorithms and compare the bounds with numerical results. Our algorithm's performance is also
studied in the presence of photon losses, where we show that it, in general, falls behind the
optimal, entanglement-based approaches. We demonstrate numerically that in this case its performance
analyzed from the Bayesian perspective coincides, in the asymptotic limit of
great resources,
with the results of a simple analysis  based on the concept of quantum Fisher
information \cite{Demkowicz2014}, where the suboptimality of such strategies had
been proved. This observation supports the recently claimed asymptotic
equivalence of Bayesian and quantum Fisher information approaches to quantum
metrology in the presence of decoherence \cite{Jarzyna2014}. Finally,
we show that while the proposed generalizations are useful from the metrological
standpoint, when considered from the point of view of Shor's algorithm, they
lose their advantage due to the different character of the relevant resource
quantification.

% Brief introduction, phase gate, bayesian approach, HS vs shot-noise. Entangled start state vs.
% "clever" phase gate application.
\section{Phase estimation}
Estimating the value of an evolution parameter based on the state of the evolved
system is a common physical task. For the purpose of this paper we consider an
estimation process consisting of the preparation of input state $\rho$;
parameter-dependent evolution $U_\varphi$, which inscribes some unknown phase
$\varphi$ onto the state ($\rho_\varphi=U_{\varphi} \rho U_{\varphi}^\dagger$);
the measurement $\{\Pi_x\}$, yielding result $x$ with probability
$p(x|\varphi)=\t{Tr}(\rho_\varphi \Pi_x)$; and, finally, the estimation part,
where the phase is estimated via the estimator function $\tildevarphi(x)$ based
on the measurement result. The goal is to come up with an estimation procedure
which allows us to infer $\varphi$ with minimal error.

In the Bayesian approach this corresponds to minimizing the mean estimation cost
\begin{equation}
  \label{eq:bayesian-general}
  \mean{C} =
    \int \d \varphi \, p(\varphi) \int \d x \, p(x|\varphi) \, C(\tildevarphi(x), \varphi),
\end{equation}
where $p(\varphi)$ is the prior describing our knowledge of the parameter before
the experiment was performed and $C(\tildevarphi, \varphi)$ is an appropriate
cost function. For a flat prior $p(\varphi)=1/2\pi$ and a phase-shift-invariant
cost function
$C(\tildevarphi, \varphi) = C(\tildevarphi - \varphi)$,
it is known that a covariant estimation-measurement strategy, where the
measurement operators are labeled with the estimated values themselves and are
expressed via a single \emph{seed} operator $\Pi$,
$\Pi_{\tildevarphi} = U_\varphi \Pi U_{\varphi}^\dagger$, is optimal \cite{Holevo1982,Chiribella2004}.
In this case Eq.~\eqref{eq:bayesian-general} simplifies to
\begin{equation}
  \label{eq:bayesian-covariant}
 \Delta^2\varphi :=  \mean{C} = \Tr \left( \Pi \int \rho_{\varphi} C(\varphi)\, \tfrac{\d \varphi}{2\pi} \right).
\end{equation}
Hence, for a given input state the optimization amounts to looking for $\Pi$ yielding the minimum of the above formula with
the constraint following from the completeness and positivity of the measurement operators:
$\int U_{\varphi} \Pi U_{\varphi}^{\dagger} \, \frac{\d \varphi}{2\pi} = \openone$
and $\Pi \geq 0$. As discussed later,
this optimization poses no serious challenge, and following the standard
reasoning from \cite{Holevo1982}, the minimal cost may be easily obtained.
In what follows we use a simple cost function, $C(\varphi) = 4 \sin^2 (\varphi/2)$, which is
common in phase estimation, as it is periodic and converges to square error for
small $\varphi$. In order to make the notation more appealing, we keep writing $\Delta^2\varphi$
instead of $\langle C\rangle$, remembering that the identification of the mean cost
function with the mean square error is strictly valid only in the regime of precise estimation.

For the moment consider a situation where $\rho$ is an arbitrary $N$ qubit state and
$U_\varphi = u_\varphi^{\otimes N}$ is a simple tensor product of single-qubit phase gates
$u_{\varphi} = e^{\ii \varphi}\ket{1}\bra{1}$, where \{$\ket{0}$, $\ket{1}$\}
constitute the single-qubit computational basis.
In this case, it is known that the optimal input state is an entangled state
belonging to the $N$-qubit fully symmetric subspace, and the corresponding
minimal cost \cite{Berry2000}
\begin{equation}
\label{eq:bayesian-optimum}
\Delta^2\varphi^{\t{opt}} = 2 \left[1- \cos\left(\frac{\pi}{N+2}\right) \right] =  \frac{\pi^2}{N^2} + O(N^{-4})
\end{equation}
has the $1/N^2$ scaling referred to as the Heisenberg scaling.
Moreover, it is known that this is the ultimate bound no matter how
$N$ unitary gates act on the qubits, even if feedback schemes are allowed \cite{Dam2007}.
Note that this bound differs by a factor $\pi^2$ from the bound derived using the quantum Fisher
information approach \cite{Giovannetti2006} and is operationally better founded,
as the explicit strategy reaching this bound is provided and requires no prior
knowledge of the value of the estimated parameter.

Still, from a practical point of view it is interesting to investigate whether the above
fundamental bound can be reached with more feasible schemes that do not require experimentally
challenging preparation of a large number of entangled particles. A simple strategy based on sending
a product input state $\rho = \ket{+}\bra{+}^{\otimes N}$, where $\ket{+} = \frac{1}{\sqrt{2}}(\ket{0} + \ket{1})$,
through the $u_\varphi^{\otimes N}$ channel yields the mean square error $\Delta^2\varphi = \frac{1}{N} + O(N^{-2})$,
which corresponds to the so-called shot-noise limit and clearly fails to approach the fundamental bound.

One can, however, consider a more general scenario parameterized by an $M$-dimensional \emph{multiplicity vector}
 $\v{m}=\{m_0,\dots,m_{M-1}\}$, $N=\sum_i m_i$, where
 the product input state of $M$ qubits $\rho=\ket{+}\bra{+}^{\otimes M}$ is subject to $U^{\v{m}}_\varphi=
 \bigotimes_{i=0}^{M-1} u_\varphi^{m_i}$, $\sum_i m_i = N$,
which consists of $N$ gates $u_\varphi$ such that $m_i$ gates act sequentially
on the $i$-th qubit (see Fig.~\ref{fig:general-kitaev}).
If, instead of the Bayesian approach, we had used the quantum Fisher information as the figure of merit,
then it is easy to see that the optimal value of the quantum Fisher information could be obtained equally well
using the $N$-qubit entangled NOON state sent through the parallel channels
$U_\varphi=u_\varphi^{\otimes N}$ as well as using a single qubit
$\rho=\ket{+}\bra{+}$ and a sequence of $N$ phase gates $U_\varphi=u_\varphi^N$
\cite{Giovannetti2006, Boixo2006, Higgins2007, Demkowicz2012, Maccone2013}. In
the Bayesian approach, however, mimicking the performance of the optimal
entanglement-based parallel strategy using an unentangled one is not that
straightforward.
%For convenience, in the following sections we rescale the estimated parameter $\varphi$ to
%lie in the $[0,1)$ instead of $[0,2\pi)$, to make it more natural to write it in the binary representation. Still, all the values of
%$\Delta^2\varphi$ the are presented on the plot correspond to the original original unscaled $\varphi$ [CHECK].

\section{Kitaev's algorithm}
\begin{figure}[t]
  \centering
  \includegraphics[scale=0.22]{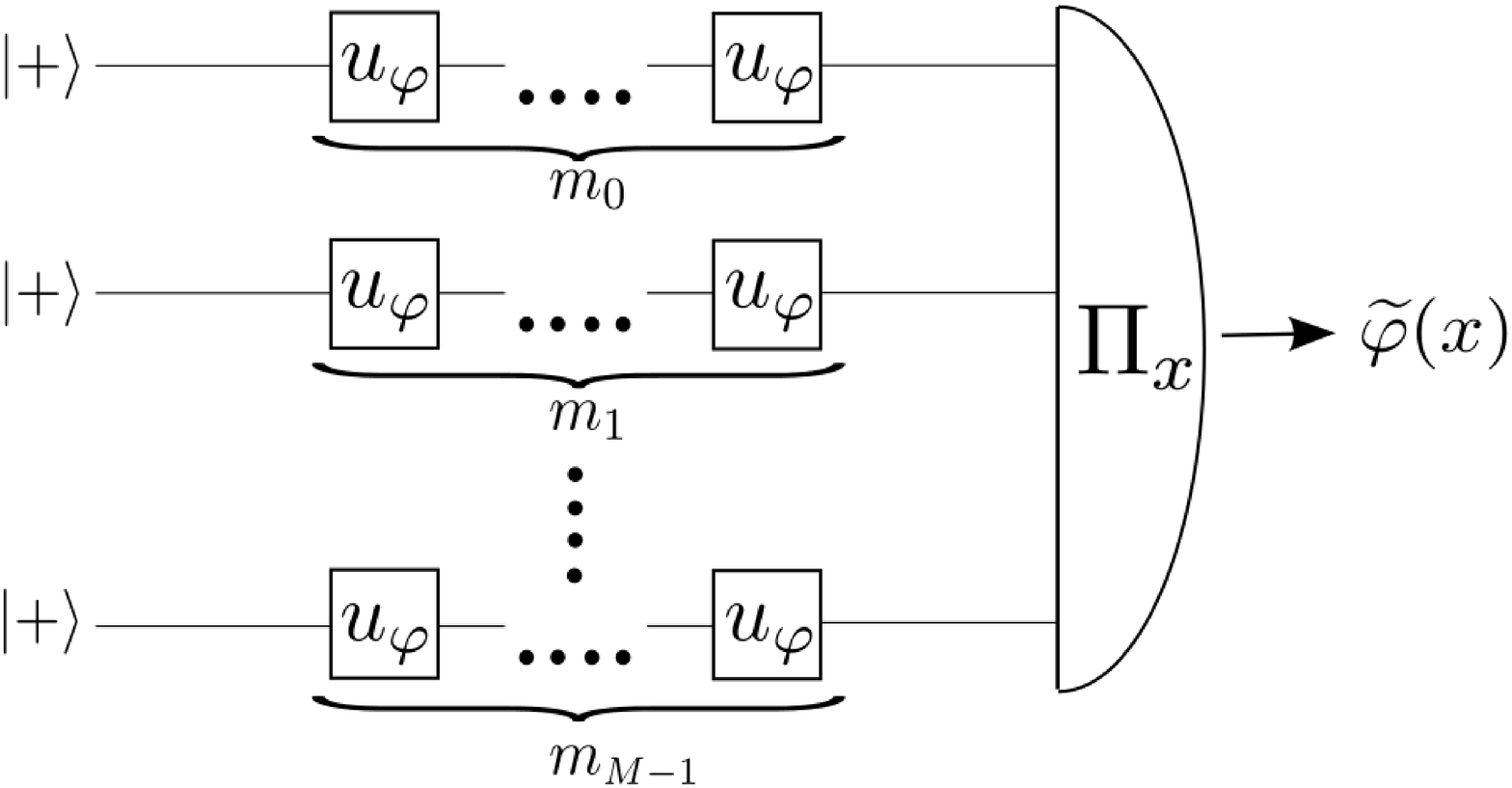}
  \caption{The circuit of a generalized Kitaev's algorithm. The $i$-th qubit
  receives the phase shift $m_i \varphi$, while the total number of used gates
  is $\sum m_i = N$. A general POVM measurement is applied and the results are
  used to construct the phase estimator.}
  \label{fig:general-kitaev}
\end{figure}

Kitaev's algorithm for phase estimation is a key building block in Shor's
factorization procedure.  Referring to the scheme in
Fig.~\ref{fig:general-kitaev} it makes use of the phase gate distribution that
corresponds to $m_i=2^{i}$, $N = 2^M-1$, while the measurement consists of
applying the inverse Fourier transform operator to the $M$ qubits and measuring
the output register in the computational basis $\ket{j}=\ket{j_0\dots j_{M-1}}$.
For further reference we denote the multiplicity vector $\v{m}$ corresponding to
the standard Kitaev's algorithm as $\v{m}_1$. The measured bits $j_i$ are then
used to obtain the estimator
$\tildevarphi = 2\pi 0.j_0\dots j_{M-1}$ \cite{Kitaev1997, Nielsen2000}.
If the underlying phase $\varphi$ is given exactly by an $M$-bit binary fraction
$\varphi=2\pi 0.j_0 j_1 \cdots j_{M-1}$, the procedure yields it correctly with
probability 1.  However, as will immediately follow from the general framework
presented in the next section, when averaged over the unknown phase
$\varphi \in [0,2\pi)$, the procedure yields the mean average cost
$\Delta^2 \varphi$, which scales as $O(N^{-1})$, even when optimized over the
measurements, and hence offers no advantage over the simple parallel strategy
using product states. This concept was also analyzed in \cite{Higgins2007} using
multiple passes of photons through wave plates, where it was shown that the
Heisenberg scaling may be regained if each of the $u_{\varphi}^{2^i}$ gates is
repeated at least $k \geq 4$ times. A theoretical foundation for this
is given in \cite{Berry2009}, where it is shown that Heisenberg scaling can be
achieved for $k \geq 3$, while setting $k = 2$ will yield an estimation cost
that scales with $O\left(\frac{\log N}{N^2} \right)$, which is already a
significant improvement over the shot-noise limit.

In the next section we present a generic extension of the Kitaev's algorithm.
The generalization together with the framework for quantifying the estimation
performance not only will allow us to analyze the setups from \cite{Berry2009}
using the general Bayesian approach, but also will take into account algorithms
where the repetition constraints are relaxed. The framework is robust in that,
for a given gate configuration, it provides the optimal end measurement and the
achieved estimation cost. Using it we analytically derive the upper bounds for
$k = 2$ and $k = 3$, which are in line with the findings in \cite{Berry2009},
and compare them with the numerically obtained optimal strategies.

% Describe algorithm, cost equation and its derivation, analytical bounds - conclusions about
% required photon repetitions. Discussion on the generality of derived results.

\section{Enhanced Kitaev's algorithm}
\label{sec:enhanced-algorithm}
For a general Kitaev-like strategy parameterized with the vector $\v{m}$, the
output state
$\ket{\psi_\varphi} = U^{\v{m}}_\varphi \ket{+}^{\otimes M}$, $\rho_\varphi = \ket{\psi_\varphi}\bra{\psi_\varphi}$,
reads explicitly
\begin{equation}
  \ket{\psi_\varphi} = \frac{1}{\sqrt{2^M}} \sum_{j=0}^{2^M-1} e^{\ii n_{\v{m}}(j) \varphi} \ket{j},
\end{equation}
where $\ket{j}=\ket{j_0\dots j_{M-1}}$ denotes a state from the computational
basis and $n_{\v{m}}(j) = \sum_i m_i j_i$ represents the multiplicity of phase
$\varphi$ that is acquired by the state. The above formula may be rewritten in
the more appealing form
\begin{equation}
  \label{eq:psivarphi}
  \ket{\psi_\varphi} = \frac{1}{\sqrt{2^M}}\sum_{n=0}^N e^{\ii n \varphi} \sqrt{J_{\v{m}}(n)} \ket{n},
\end{equation}

where $J_{\v{m}}(n) = \sum_{j=0}^{2^M-1} \delta_{n,n_{\v{m}}(j)}$ corresponds to
the number of basis vectors that evolve with a given phase multiplicity $n$,
while
\begin{equation}
  \ket{n} = \frac{1}{\sqrt{J_{\v{m}}(n)}}\sum_{j=0}^{2^M-1} \delta_{n,n_{\v{m}}(j)} \ket{j}
\end{equation}
is the normalized equally weighted superposition of these vectors.
Plugging Eq.~\eqref{eq:psivarphi} into Eq.~\eqref{eq:bayesian-covariant}
and performing the integration we arrive at the formula for the average cost,
\begin{equation}
\label{eq:costpi0}
\begin{split}
  \Delta^2\varphi_{\v{m}} &= \frac{1}{2^{M-1}}\Big(\sum_{n=0}^N J_{\v{m}}(n) \Pi_{n,n} + \\
      &-\sum_{n=0}^{N-1} \Pi_{n,n+1} \sqrt{J_{\v{m}}(n)J_{\v{m}}(n+1)}
\Big),
\end{split}
\end{equation}
where $\Pi_{n,m}= \bra{n} \Pi \ket{m}$. The completeness condition
$\int U^{\v{m}}_\varphi  \Pi U^{\v{m}\dagger}_\varphi \frac{\t{d}\varphi}{2\pi} = \openone$
implies $\Pi_{n,n}=1$, while the positivity constraint
$\Pi \geq 0$ implies that $|\Pi_{n,m}| \leq \sqrt{\Pi_{n,n} \Pi_{m,m}} =1$.
Hence, the optimal choice for the seed operator $\Pi$, which minimizes $\Delta^2\varphi$ in
Eq.~\eqref{eq:costpi0}, corresponds to the choice $\Pi_{n,m}=1$. Vectors
$\{\ket{n}\}$ do not, in general, span the whole $2^M$-dimensional Hilbert
space, so the explicit form of the optimal seed operator $\Pi$ reads
\begin{equation}
\Pi = \sum_{n, m}\ket{n}\bra{m} + \openone_{\bot},
\end{equation}
where $\openone_{\bot}$ represents the identity operator on the subspace
orthogonal to the one spanned by $\{ \ket{n} \}$. This allows us to produce the
set of POVMs $\{\Pi_{\varphi}\}$ parameterized by the continuous index
$\varphi$, which, if required, can also be replaced with a finite number of
operators \cite{Derka1997, Dam2007}. The resulting minimal cost reads
\begin{equation}
  \label{eq:algorithm-cost}
  \Delta^2\varphi_{\v{m}} = 2 - \frac{1}{2^{M-1}} \sum\limits_{n=0}^{N-1}
    \sqrt{J_{\v{m}}(n) J_{\v{m}}(n+1)}.
  \end{equation}
From the above formula it readily follows that for the standard Kitaev's
algorithm, parameterized by $\v{m} = \v{m}_1$ where $m_i=2^{i}$, we have
 $N=2^M-1$ and $J_{\v{m}_1}(n)=1$, which results in
\begin{equation}
  \Delta^2\varphi_{\v{m}_1} = 2- \frac{2^M -1}{2^{M-1}} = \frac{1}{2^{M-1}} = \frac{2}{N+1} .
\end{equation}
Hence, as mentioned before, the strategy offers no advantage over the basic
parallel product-state-based strategy.

Let us now focus on two more interesting cases,
$\v{m}_2 = \v{m}_1 \wedge \v{m}_1$ and
$\v{m}_3 = \v{m}_1 \wedge  \v{m}_1 \wedge \v{m}_1$, where $\wedge$ denotes
vector concatenation. The cases are extensions of the original Kitaev's
algorithm, where each of the $u_{\varphi}^{2^i}$ phase shifts is applied to two
or three different qubits, respectively. Using Eq.~\eqref{eq:algorithm-cost} it
is possible to obtain analytical bounds on the achievable precision in terms of
the total consumed resources $N$ for these two cases, whose derivations are
described in more detail in the Appendix, and read
\begin{equation}
  \label{eq:double-repetition-cost}
  \Delta^2 \varphi_{\v{m}_2} \leq 4 \frac{\ln(N+2) - \ln 2 + 3}{(N+2)^2}
    = O \left( \frac{\log N}{N^2} \right),
\end{equation}
\begin{equation}
  \label{eq:triple-repetition-cost}
  \Delta^2 \varphi_{\v{m}_3} \leq \frac{27}{(N+3)^2} \frac{N+1}{N+3}
    = O \left( \frac{1}{N^2} \right).
\end{equation}
This shows that our estimation costs in these two simple extensions of Kitaev's
algorithm are asymptotically equivalent to those presented in
\cite{Berry2009}.

% Describe numerical approaches, provide juxtaposition plots.
\section{Cost analysis}
\label{sec:cost-analysis}
The upper bounds in the previous section were derived using simple inequalities
between means, which raises the question how much more efficient the
estimation costs actually are. It would also be insightful to find out how much
improvement can be gained by allowing more general gate configurations. In order
to address these problems to some extent, numerical simulations for tractable
space sizes were carried out. For a given $\v{m}$, the corresponding
$J_{\v{m}}(n)$ were calculated, which in turn allowed us to compute the
estimation cost using (\ref{eq:algorithm-cost}). In order to cut down on
computation time, only a subspace of nondecreasing vectors $\v{m}$ was
considered. We constrained the problem by allowing only powers of 2 as
multiplicities. We looked into cases where $M \leq 32$. To further cut down on
computation time we enforced at least two repetitions of each multiplicity for
$M \in [21, 25]$ and at least three for $M \in [26, 32]$. The outcomes were then
aggregated with respect to $N = \sum m_i$ and the lowest cost was stored as the
best result found.

\begin{figure}[t]
  \centering
  \includegraphics[scale=0.288]{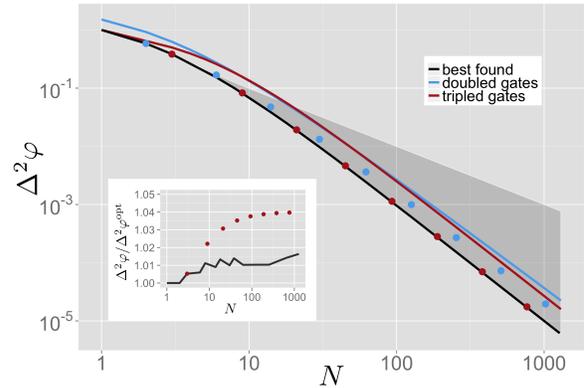}
  \caption{Plot showing the estimation costs with respect to the number of used
  resources. The black line connects the best numerically found solutions.
  Filled blue (red) circles show the results for $\v{m}_2$ ($\v{m}_3$), while
  lines are the corresponding analytical upper bounds. The dark-gray ribbon
  depicts the area between the Bayesian estimation optimum and the shot-noise
  limit given by $\v{m}_{\otimes}$. Inset: Cost ratios of the best found
  parameter $\v{m}$ to the Bayesian optimum. Ratios correspond to the costs
  presented in the main plot.}
  \label{fig:cost-plot}
\end{figure}

The results are presented in Fig.~\ref{fig:cost-plot}. They are compared with
the optimal cost given by \eqref{eq:bayesian-optimum} and a shot-noise benchmark
obtained using $\v{m} = \v{m}_{\otimes}$, where $m_i = 1$, which mimics a
classical experiment with repeated estimation using single passes through a
$u_{\varphi}$ gate. The numerically obtained cost values for resource numbers
$N$ are indeed visibly lower than the analytical bounds, although the asymptotic
behavior seems to be preserved.
The costs found by the algorithm are nearly optimal. Even though the general
structure of the best $\v{m}$ for an arbitrary $N$ seems contrived, for
$\v{m}_3$ (filled red circles) the costs are already barely discernible from the
Bayesian optimum, which is a new finding. For $\v{m}_2$ the computed cost is
lower than the corresponding bound, yet visibly worse than the optimum. Under
the above conditions we only searched through all viable $\v{m}$ values for
$N \leq 20$, however, we obtained consistent results for up to $N =1000$
resources. As shown in the inset in Fig.~\ref{fig:cost-plot} the found results
are worse than the optimum by only less than $2\%$. The slightly saw-like shape
of the ratios stems from the constraints on $\v{m}$. The value of
$\Delta^2 \varphi / \Delta^2\varphi^{\t{opt}}$ for $\v{m}_3$ was also computed
for $N \leq 10^5$ and it seems to converge around $1.04$. Our algorithm
enhancement can thus deliver estimation costs within a very few percent of the
optimum.

% Discuss cases with losses. Provide plots and conclusions.
\section{Setup with photon losses}
The efficiency of quantum measurements and algorithms is highly dependent on
external noise factors. In this article we focus on one particular noise model,
namely, photon losses. The topic has been considered in previous works
\cite{Kolodynski2010,Knysh2010,Demkowicz2010,Escher2011,Demkowicz2014,Jarzyna2014}
and optimal estimation bounds for the most general strategies are known. Here we
want to investigate the ultimate performance of Kitaev-like protocols in the
presence of losses from the Bayesian perspective. In our analysis we assume that
each phase shift gate has probability $\eta$ of functioning correctly. For the
repeated $u_{\varphi}^m$ gates the success probability decreases exponentially
to $\eta^m$ and the chances of losing the photon become $1 - \eta^m$. This is
equivalent to modeling $u_\varphi$ as a Mach-Zender interferometer with phase
delay $\varphi$ and with power transmission equal to $\eta$ in both arms.

It turns out that we can present the costs for the lossy scenario using
noiseless cost values. This follows directly from the formula
\begin{equation}
  \label{eq:lossy-cost-formula}
  \Delta^2\varphi_{\v{m}, \eta} = \displaystyle\sum\limits_{\v{n} \in \{0,1\}^M}
    \Delta^2 \varphi_{\v{m} \v{n}} \prod_{i=0}^{M-1} \eta^{m_i n_i}
    (1-\eta^{m_i})^{1-n_i},
\end{equation}
where vector multiplication is element-wise. The cost is simply a weighted
average of noiseless costs where the parameter vectors $\v{m}\v{n}$ are obtained
by removing dissipated photons from $\v{m}$. Note that a faulty gate eradicates
all information about the phase contained in a particular qubit. By intuition it
seems natural to expect that the stronger the noise factor, the lower the phase
gate multiplicities of optimal circuits will be.

Simulating Eq.~\eqref{eq:lossy-cost-formula} would be computationally
infeasible for $N >> 10$; instead we propose the following approximation. Each
$u_{\varphi}^m$ gate is a Bernoulli trial with a probability of success equal to
$\eta^m$. Repeating it $\eta^{-m}$ times will, on average, yield a successful
phase windup, as it is the expected number of trials until first success. This
translates to a modified resource expense of $N = \sum_i \eta^{-m_i} m_i$.
The estimation variance is naturally not exact for a single experiment, but it
holds in the regime of numerous repetitions. Note that losing a photon is
equivalent to its entering an orthogonal vacuum state, hence we always have
information, whether a trial ends in success or failure. The repetitions might
require fresh resources; i.e., in the multipass setting additional photons would
be required.
\begin{figure}[t]
  \centering
  \includegraphics[scale=0.07]{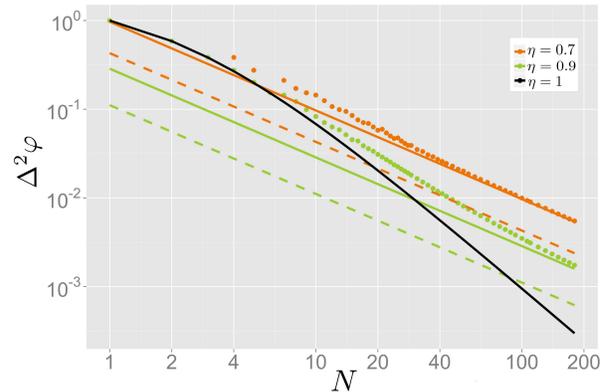}
  \caption{Plot showing the approximate performance of the enhanced Kitaev's
  algorithm for two lossy setups. Filled circles reveal the numerically computed
  cost values of the algorithm, solid lines are the cost limits for
  entanglement-free strategies, and dashed lines depict the ultimate bounds
  for the lossy setup where any entangled start states are permitted.}
  \label{fig:lossy-cost-plot}
\end{figure}
Figure~\ref{fig:lossy-cost-plot} presents the results of numerical simulations for
photon losses. Two transmission factors have been considered; the Bayesian
noiseless optimum given by Eq.~\eqref{eq:bayesian-optimum} is juxtaposed for
comparison. Filled circles depict the numerically found estimation costs, using the
same algorithm as in Sec.~\ref{sec:cost-analysis}, yet with updated resource
expenditure. They visibly converge to the solid lines given by
\begin{equation}
  \Delta^2 \varphi_{\eta}^{\mathrm{unentangled}} = \frac{e \ln \left( 1 / \eta \right) }{N},
\end{equation}
which represent the asymptotic optimal cost obtained within the quantum Fisher information approach  for unentangled systems \cite{Demkowicz2014}.
The convergence of the numerical Bayesian results to the asymptotic cost obtained within the quantum Fisher information approach
should be viewed as an illustration of a general claim of asymptotic equivalence of the two approaches in the case where
asymptotic scaling of the mean square error is limited to $O(N^{-1})$ \cite{Jarzyna2014}.
The ultimate bounds (dashed lines), which are valid for any input states including
highly entangled ones, are equal to \cite{Kolodynski2010, Knysh2010, Escher2011, Demkowicz2012} 
\begin{equation}
  \Delta^2 \varphi_{\eta}^{\mathrm{general}} = \frac{1 - \eta}{\eta N}
\end{equation}
and are clearly out of reach of the
presented strategy. This proves that in the presence of losses, unlike in the decoherence-free case,  entanglement is essential 
for reaching the ultimate precision bound---a fact demonstrated before only within the quantum Fisher information approach 
\cite{Demkowicz2010, Demkowicz2014}.

% Describe different resource interpretations, brief intro to Shor's algorithm and ideas
% why the improvement wouldn't help.
\section{Quantum resources and Shor's algorithm}
The phase estimation procedure is a crucial part of Shor's algorithm \cite{Shor1997}, hence this
article would be incomplete if no reference to the famous factorization
algorithm were made. Unfortunately, as promptly shown, the proposed enhancements
to the phase-estimation algorithm, while useful from the metrological
perspective, are of little use from the computational point of view. We do not
discuss the whole of Shor's algorithm, as it is available in the original
article and in other works \cite{Nielsen2000, Ekert1996} with detailed examples;
we only touch on the parts of the procedure which directly relate to our
subject.

The core of Shor's algorithm is the subroutine for \emph{order finding}. The task is as follows:
given integers $x$ and $N$, $x < N$, we are to find the smallest positive integer $r$ such that
$x^r \equiv 1\  (\t{mod } N)$. This problem is equivalent to factorizing $N$
in the sense that, with an additional expense of polynomial (classical)
computation time, we can obtain the solution of one using the other. To embed
order finding in the phase estimation setting we utilize the following gate:
$u_x \ket{y} = \ket{xy \mod N}$. Regardless of the value of $x$ the operator
has eigenstates $\ket{e_s}, s \in \{0, 1, \dots, r-1\}$, where
\begin{equation}
  \label{eq:shor-eigenstates}
  u_x \ket{e_s} = e^{\frac{\ii 2 \pi s}{r}} \ket{e_s}.
\end{equation}
By employing Kitaev's algorithm with $u_x$ as the phase imprinting gate (with
$x$ selected randomly) and further quantum-mechanical and number-theoretic
transformations the algorithm produces $r$ with a high probability.

The intrinsic structure of operator $u_x$ underpins the computational power of the algorithm in
that we have $u_{x}^{m} = u_{x^m \t{ mod } N}$. Hence the sequences of controlled $u_{x}^{2^i}$ gates
can be implemented in the time polynomial with respect to $L = \lceil \log(N) \rceil$ using modular
exponentiation. We juxtapose this case with general phase estimation as discussed in earlier
sections. Previously we assumed the expense of utilizing a gate $u_\varphi^m$ to be
$m$ (e.g., $m$ photon passes). Here, operating gate $u_{x}^{m}$ will still cost us 1,
plus the additional computation time to obtain $x^m \mod N$. Using the standard ``elementary''
multiplication algorithm and modular exponentiation we can perform this computation in
time $O \left( L^2 \log m \right)$, which can also be bounded by $O\left(L^3 \right)$. We see that
in this case a gate $u_x^m$ can be implemented with a single resource unit, regardless of $m$, and at the
cost of computation, whose time is bounded by a constant for a set value of $L$. The approach with
repeating gates with lower phase multiplication factors would be counter-productive here.

To conclude this section we remark that when we equate physical resources with
the number of qubits $M$ involved, irrespective of the multiplicity of the phase
shifts they experience, that is, set $N := M$, then the original Kitaev's
algorithm parameterized by $\v{m}_1$ is indeed the optimal one in terms of cost
in the Bayesian approach, and it is also the only optimal solution. A proof of this statement is
easily obtained as follows. We assume without loss of generality that $m_i > 0$. Then
$J_{\v{m}}(0) = J_{\v{m}}(N) = 1$, and again applying the inequality between the
arithmetic and the geometric mean to Eq.~\eqref{eq:algorithm-cost} and using
$\sum_n J_{\v{m}}(n) = 2^M$, we obtain
$\forall_{\v{m}}\Delta^2 \varphi_{\v{m}} \geq \Delta^2 \varphi_{\v{m}_1}$. We
have equality iff $J_{\v{m}}(n) = 1$, which is only true when $\v{m} = \v{m}_1$.

\section{Conclusions}
We have found that Kitaev's phase-estimation algorithm can be naturally
generalized to reach optimality in the sense of the Bayesian estimation theory.
The enhancement is based on repeated applications of the phase gates and does
not require entanglement, although it makes use of general measurements, whose
experimental implementation may still pose a challenge. We introduced a
framework for analyzing generalizations of Kitaev's algorithm with the
ready-to-use recipe for calculating their costs. We then rederived the
analytical bounds first presented in \cite{Berry2009} for Kitaev's algorithm
with doubled and tripled gates. For moderate resource numbers we found
algorithm enhancements which approach closely the Bayesian estimation optimum.
The behavior of our algorithm was also analyzed in a setup with photon losses
where it was shown that they cannot match the performance of the optimal
entanglement-based strategies. We have also confirmed the asymptotic equivalence
of Bayesian and quantum Fisher information approaches in the problem analyzed.
Finally, we considered the enhancement from the point of view of Shor's
factorization procedure. A negative conclusion was reached as to its usefulness
for this purpose and the original Kitaev's algorithm was found to be the only
optimal solution, thus revealing an interesting side to the understanding of
quantum resources and their utilization in various measurement schemes,
including quantum algorithms.

\begin{acknowledgments}
This research work was supported by the European Commission Seventh Framework
Programme project SIQS (Simulators and Interfaces with Quantum Systems)
cofinanced by the Polish Ministry of Science and Higher Education.
\end{acknowledgments}

%Appendices
\appendix

\begin{widetext}

\section*{Appendix: Derivation of estimation bounds for doubled and tripled gates}
\label{apx:positional-systems}
In this section we give examples of how Eq.~\eqref{eq:algorithm-cost} can be
used to obtain costs for various generalizations of Kitaev's algorithm. Earlier
we calculated the cost for $\v{m}_1$; here we investigate the cases of $\v{m}_2$
and $\v{m}_3$, which are of interest from the point of view of this article.
Before we do so, however, we take a brief look at what happens when
$\v{m} = \v{m}_{\otimes}$; this corresponds to the tensor product of
single-qubit phase shift gates, which we have used as a ``classical'' benchmark
before. Using our framework and noting that
$J_{\v{m}_{\otimes}}(n) = \binom{N}{n}$ we obtain the cost
$\Delta^2 \varphi_{\v{m}_{\otimes}}$, which, as expected, can be shown to
approach $N^{-1}$ for large $N$.

The key to making practical use of Eq.~\eqref{eq:algorithm-cost} is finding the structure of
$J_{\v{m}}(n)$. It is the number of vectors evolving at phase multiplicity $n \varphi$ and also the
number of representations of $n$ in a positional system given by $\v{m}$. For $\v{m}_2$ it is a
``pseudobinary'' system where each position is repeated twice. The function can be written in the
following recursive form for $0 \leq n < 2^{\frac{M}{2}}$:
\begin{equation}
 J_{\v{m}_2}(n) = \left\{ \begin{array}{ll}
   J_{\v{m}_2}\left(\frac{n}{2}\right) + J_{\v{m}_2}\left(\frac{n-2}{2}\right) & : 2 \mid n \\
   2 J_{\v{m}_2}\left(\frac{n-1}{2}\right) & : 2 \centernot \mid n
 \end{array} \right. .
\end{equation}
For even $n$ all valid representations either have 0's at the two least significant
bits (of our modified positional system) or have them both set to 1 and use the remainder to
represent $n - 2$. If $n$ is odd, we have to set exactly one of the two
least significant bits, while the rest represent $n-1$. In both cases we can
equivalently discard the two fixed bits and look at values shifted to the right
by two positions, which is the same as dividing the value by 2. By further
noting that $J_{\v{m}_2}(0) = 1$ we get $J_{\v{m}_2}(n) = n + 1$ for
$0 \leq n < 2^{\frac{M}{2}}$. On the other hand, for
$2^{\frac{M}{2}} \leq n < 2^{\frac{M}{2}+1} - 1$ the case is analogous when we
swap 0's and 1's, and we then get
$J_{\v{m}_2}(n) = J_{\v{m}_2} \left( 2^{\frac{M}{2} + 1} - n - 2 \right)$.

We can now present the cost formula for $\v{m}_2$
\begin{equation}
 \Delta^2\varphi_{\v{m}_2} = 2 - \frac{1}{2^{M-2}} \sum_{n=0}^{2^{\frac{M}{2}} - 1}
   \sqrt{n(n + 1)} .
\end{equation}
The sum can be bounded using the inequality between the geometric and the
harmonic mean,
\begin{equation}
 \sum\limits_{n=0}^{2^{\frac{M}{2}} - 1} \sqrt{n(n + 1)}
     \geq \sum\limits_{n=0}^{2^{\frac{M}{2}} - 1} \left( n + \frac{n}{2n + 1} \right)
     = 2^{M - 1} - \frac{1}{2} \sum\limits_{n=0}^{2^{\frac{M}{2}} - 1} \frac{1}{2n + 1}
     \geq 2^{M -1} - \frac{1}{4}H(2^{{\frac{M}{2}}}) - \frac{1}{2},
\end{equation}
where $H(n)$ is the $n$th harmonic number. Using $H(n) \leq \ln n + 1$ and
setting $N := 2 \left( 2^{\frac{M}{2}} - 1 \right)$, we get the bound from
Sec.~\ref{sec:enhanced-algorithm}.

For $\v{m}_3$ the derivation is analogous, although considerably more laborious.
The formula, Eq.~\eqref{eq:algorithm-cost}, reads
\begin{equation}
  \Delta^2\varphi_{\v{m}_3} = 2 - \frac{1}{2^{M-1}} S_{\v{m}_3} ,
\end{equation}
where
\begin{equation}
  \begin{split}
  S_{\v{m}_3} &=
      \sum\limits_{k=1}^{2^{{\frac{M}{3}} - 1}} (k+1)\sqrt{k(k+2)} +
      \frac{3}{4} 2^{{\frac{2M}{3}}} +
      \frac{3}{2} \sum\limits_{k=1}^{2^{{\frac{M}{3}} - 1} - 1}
        \sqrt{\left(2^{\frac{2M}{3}} - \frac{4k(k-1)}{3} \right) \left(2^{\frac{2M}{3}} - \frac{4k(k+1)}{3} \right)} \\
    &\geq
      2^M - 2^{\frac{M}{3}} - \frac{8}{3}\sum\limits_{k=1}^{2^{\frac{M}{3} - 1} - 1} \frac{k^2}{2^{\frac{2M}{3}} - \frac{4 k^2}{3}}
    \geq 2^M - \frac{3}{2} 2^{\frac{M}{3}} + 1.
  \end{split}
\end{equation}
Combined with $N := 3 \left( 2^{\frac{M}{3}} - 1 \right)$ the above is then easily transformed to
Eq.~\eqref{eq:triple-repetition-cost}.
\end{widetext}

\bibliographystyle{apsrev4-1}

\end{document}